\documentstyle[12pt]{article}
\def\thebibliography#1{\section*{
\begin{center}
References
\end{center}
}\list
 {\arabic{enumi}) }{\settowidth\labelwidth{[#1]}\leftmargin\labelwidth
 \advance\leftmargin\labelsep
 \usecounter{enumi}}
 \def\newblock{\hskip .11em plus .33em minus .07em}
 \sloppy\clubpenalty4000\widowpenalty4000
 \sfcode`\.=1000\relax}

\renewcommand{\theequation}{\mbox{\arabic{section}$\cdot$\arabic{equation}}}

\scrollmode
\begin{document}
\title{Generators of Internal Lorentz Transformations and of General
Affine Coordinate Transformations in Teleparallel Theory of
(2+1)-Dimensional Gravity\\--- Cases with Static Circularly Symmetric
Space-Times---}
\author{Toshiharu K{\footnotesize AWAI}
\\{\normalsize \em Department of Physics, Osaka City University, Sugimoto,
Sumiyoshi-ku, Osaka 558}}
\date{}
\maketitle

\begin{abstract}
In a teleparallel theory of (2+1)-dimensional gravity developed
in a previous paper, we examine generators of internal Lorentz
transformations and of general affine coordinate transformations
for static circularly symmetric exact solutions of gravitational
field equation. The \lq \lq spin" angular momentum, the
energy-momentum and the \lq \lq extended orbital angular momentum"
are explicitly given for each solution. Also, we give a critical
comment on Deser's claim that neither momentum nor boosts are
definable for finite energy solutions of three-dimensional Einstein
gravity.
\end{abstract}

\section*{
\begin{center}
\S~1. Introduction
\end{center}
}
\addtocounter{section}{1}
                                    \setcounter{equation}{0}

\mbox{\hspace*{3ex}}In a previous $\mbox{\rm paper},^{1)}$ we have
developed a teleparallel theory of (2+1)-dimensional gravity,
in which three types of static circularly symmetric exact solutions of
the gravitational field equation are given, and horizons and
singularities of space-times given by these solutions have been examined.
Two of these solutions are also solutions of the three-dimensional
Einstein equation.

In the Einstein theory of three-dimensional
$\mbox{\rm gravity}^{2),3)}$ and in the theory of (2+1)-dimensional
$\mbox{\rm supergravity},^{4)}$ global charges such as energy-momentum,
angular momentum and supercharge have been examined, and the following
has been claimed: (1)Neither momentum nor boosts are definable
for finite energy solutions in the three-dimensional Einstein
$\mbox{\rm theory}.^{2)}$ (2)There is no way to define linear momentum and
supercharge for a generic solution to (2+1)-dimensional
$\mbox{\rm supergravity}.^{4)}$

Thus, it would be significant to examine global charges in the
teleparallel theory of (2+1)-dimensional gravity. The main purpose of
this paper is to investigate the generators of the internal
Lorentz transformations and of the general affine coordinate
transformations for the exact solutions mentioned above.
\section*{
\begin{center}
\S~2. Basic Framework of the Theory and Generalities
on Generators
\end{center}
}
\addtocounter{section}{1}
                                       \setcounter{equation}{0}

\mbox{\hspace*{3ex}}For the convenience of the latter discussion, we
briefly summarize the basic part of the teleparallel theory developed in
Ref. 1).

The three-dimensional space-time is assumed to be a differentiable
manifold endowed with a Lorentzian metric
$g_{\mu \nu }dx^{\mu }\otimes dx^{\nu }$ related to the fields
${\bf e}^{k}={e^{k}}_{\mu }dx^{\mu }\; (k=0, 1, 2)$ through the
relation $g_{\mu \nu }={e^{k}}_{\mu }\eta_{kl}{e^{l}}_{\nu }$ with
$(\eta_{kl}) \stackrel{\rm def}{=}$diag$(-1, 1, 1)$. Here,
$\{x^{\mu }\:;\mu=0, 1, 2\}$ is a local coordinate of the space-time.
The fields ${\bf e}_{k}={e^{\mu }}_{k}\partial /\partial x^{\mu }$,
which are dual to ${\bf e}^{k}$, are the dreibein fields.
The strength of ${e^{k}}_{\mu }$ is given by ${T^{k}}_{\mu \nu }
=\partial_{\mu }{e^{k}}_{\nu }-\partial_{\nu }{e^{k}}_{\mu }$.
The covariant derivative of the Lorentzian vector field $V^{k}$ is
defined by $\nabla_{l}V^{k}\stackrel{\rm def}{=}
{e^{\mu }}_{l}\partial_{\mu }V^{k}$,
and the covariant derivative of the world vector field ${\bf V}=
V^{\mu }\partial /\partial x^{\mu }$ with respect to the affine
connection $\Gamma^{\mu }_{\lambda \nu }$ is given by
$\nabla_{\nu }V^{\mu }=\partial_{\nu }V^{\mu }
+\Gamma^{\mu}_{\lambda \nu }V^{\lambda }$.
The requirement $\nabla_{l}V^{k}=
{e^{\nu }}_{l}{e^{k}}_{\mu }\nabla_{\nu }V^{\mu }$
for $V^{\mu } \stackrel{\rm def}{=}{e^{\mu }}_{k}V^{k}$ leads to
\begin{eqnarray}
{T^{k}}_{\mu \nu }
&\equiv &{e^{k}}_{\lambda }{T^{\lambda }}_{\mu \nu }\;,\\
{R^{\mu }}_{\nu \lambda \rho } &\stackrel{\rm def}{=}&
\partial_{\lambda }\Gamma^{\mu }_{\nu \rho }-
\partial_{\rho }\Gamma^{\mu }_{\nu \lambda }+
\Gamma^{\mu }_{\tau \lambda }\Gamma^{\tau }_{\nu \rho }-
\Gamma^{\mu }_{\tau \rho }\Gamma^{\tau }_{\nu \lambda }\equiv 0\;,\\
\nabla_{\lambda }g_{\mu \nu } &\stackrel{\rm def}{=}&
\partial_{\lambda }g_{\mu \nu }-
\Gamma^{\rho }_{\mu \lambda }g_{\rho \nu }-
\Gamma^{\rho }_{\nu \lambda }g_{\rho \mu }\equiv 0\;,
\end{eqnarray}
where ${T^{\lambda }}_{\mu \nu }$ is defined by
${T^{\lambda }}_{\mu \nu }\stackrel{\rm def}{=}
\Gamma^{\lambda }_{\nu \mu }-\Gamma^{\lambda }_{\mu \nu }$.
The components ${T^{\lambda }}_{\mu \nu }$ and
${R^{\mu }}_{\nu \lambda \rho }$ are those of the torsion tensor and
of the curvature tensor, respectively. Equation (2$\cdot $2)
implies the teleparallelism. The field components ${e^{k}}_{\mu }$ and
${e^{\mu }}_{k}$ are used to convert Latin and Greek indices. Also,
raising and lowering the indices $k, l, m, ...$ are accomplished
with the aid of $(\eta^{kl})=(\eta_{kl})^{-1}$ and $(\eta_{kl})$,
respectively.

For the matter field $\varphi $ belonging to a representation of the
three-dimensional Lorentz group,
$\L_{M}(\varphi , \nabla_{k}\varphi )$ with $\nabla_{k}\varphi
\stackrel{\rm def}{=}{e^{\mu }}_{k}\partial_{\mu }\varphi $ is a
Lagrangian density invariant under global Lorentz transformations
and under general coordinate transformations, if
$L_{M}(\varphi ,\partial_{k}\varphi )$ is an invariant Lagrangian
density on the three-dimensional Minkowski space-time. For the
dreibein fields ${\bf e}_{k}$, we have $\mbox{\rm employed}^{1)}$
\begin{equation}
L_{G}=c_{1}t^{klm}t_{klm}+c_{2}v^{k}v_{k}+c_{3}a^{klm}a_{klm}\;,
\end{equation}
as the Lagrangian density. Here, $t_{klm}\,, v_{k}\,, $ and $a_{klm}$
are the irreducible components of $T_{klm}$, which are defined by
\begin{eqnarray}
 t_{klm} & \stackrel{\rm def}{=} & \frac{1}{2}(T_{klm}+T_{lkm})
 +\frac{1}{4}(\eta_{mk}v_{l}+\eta_{ml}v_{k})
 -\frac{1}{2}\eta_{kl}v_{m}\;, \nonumber \\
     & & \\
 v_{k} & \stackrel{\rm def}{=} & {T^{l}}_{lk}\;,
\end{eqnarray}
and
\begin{equation}
a_{klm}\stackrel{\rm def}{=}\frac{1}{3}(T_{klm}+T_{mkl}+T_{lmk})\;,
\end{equation}
respectively, and $c_{1}\,,c_{2}$ and $c_{3}$ are real constant
parameters.$^{*)}$ Then,
\begin{equation}
{\bf I}\stackrel{\rm def}{=}\frac{1}{c}\int {\bf L}dx^{0}dx^{1}dx^{2}
\end{equation}
-----------------------------------------------------------------------\\
$^{*)}$ The parameters $c_{1}\, , c_{2}$ and $c_{3}$ correspond to
$\alpha \, ,\beta $ and $\gamma $ in Ref. 1), respectively.\\
is the total action of the system, where $c$ is the light velocity
in vacuum and ${\bf L}$ is defined by ${\bf L} \stackrel{\rm def}{=}
{\bf L}_{G}+{\bf L}_{M}$ with ${\bf L}_{G} \stackrel{\rm def}{=}
\sqrt{-g}L_{G}\,, {\bf L}_{M} \stackrel{\rm def}{=}
\sqrt{-g}L_{M}(\varphi , \nabla_{k}\varphi )$ and
$g \stackrel{\rm def}{=} \det(g_{\mu \nu })$.

The gravitational field equation following from the action {\bf I} is
\begin{equation}
-2\nabla^{m}F_{klm}+2v^{m}F_{klm}+2H_{kl}-\eta_{kl}L_{G}=T_{kl}
\end{equation}
with
\begin{eqnarray}
F_{klm}& \stackrel{\rm def}{=} &c_{1}(t_{klm}-t_{kml})+
c_{2}(\eta_{kl} v_{m}-\eta_{km}v_{l})+2c_{3}a_{klm}
=-F_{kml}\;, \\
H_{kl}& \stackrel{\rm def}{=} &T_{mnk}{F^{mn}}_{l}-
\frac{1}{2}T_{lmn}{F_{k}}^{mn}=H_{lk}\;,\\
\nabla^{m}F_{klm}& \stackrel{\rm def}{=} &
e^{\mu m}\partial_{\mu }F_{klm}\;.
\end{eqnarray}
Also, $T_{kl}$ is the energy-momentum density defined by
\begin{equation}
\sqrt{-g}T_{kl} \stackrel{\rm def}{=} e_{l\mu }
\frac{\delta {\bf L}_{M}}{\delta {e^{k}}_{\mu }}
\stackrel{\rm def}{=}
e_{l\mu }\left\{\frac{{\bf L}_{M}}{\partial {e^{k}}_{\mu}}
-\partial_{\nu }\left(\frac{\partial{\bf L}_{M}}
{\partial({\partial_{\nu }e^{k}}_{\mu })}\right)\right\}\;.
\end{equation}
The field equation (2$\cdot $9) reduces to the three-dimensional
Einstein $\mbox{\rm equation},^{1)}$ when the conditions
\begin{equation}
3c_{1}=-4c_{2}=-8c_{3}=-\frac{1}{\kappa}\;,\; \; T_{kl}=T_{lk}\;,
\end{equation}
are satisfied. Here, $\kappa $ stands for the \lq \lq Einstein
gravitational constant" $\kappa =8\pi G/c^{4}$ with $G$ being the
\lq \lq Newton gravitational constant."

We consider the global internal Lorentz transformation
\begin{equation}
{e'^{k}}_{\mu }={e^{k}}_{\mu }
+{d^{k}}_{l}{e^{l}}_{\mu }\;,
\; \; \phi'=\phi +\frac{i}{2}d^{kl}M_{kl}\phi
\end{equation}
with $d^{kl}=-d^{lk}$ and $\{M_{kl}=-M_{lk}\,|\, k\,,l=0\,,1\,,2\}$
being an arbitrary infinitesimal constant and a set of representation
matrices of the standard basis of the Lie algebra of the
three-dimensional homogeneous Lorentz group, respectively. For the
space-like surface $\sigma $, the generator $G(\sigma )$ of this
transformation is given by
\begin{equation}
G(\sigma )=-\frac{1}{2}d^{kl}S_{kl}
\end{equation}
with the \lq \lq spin" angular $\mbox{\rm momentum}^{*)}$
\vspace*{0.5 cm}\\
-----------------------------------------------------------------------\\
$^{*)}$ By \lq \lq spin" we mean the quantum number
associated with the three-dimensional Lorentz group.
                                          \pagebreak
\begin{equation}
S_{kl} \stackrel{\rm def}{=} \int_{\sigma }
{{\bf S}_{kl}}^{\mu }d\sigma_{\mu }\;.
\end{equation}
Here, ${{\bf S}_{kl}}^{\mu }$ is defined $\mbox{\rm by}^{*)}$
\begin{equation}
{{\bf S}_{kl}}^{\mu } \stackrel{\rm def}{=}
-2{{\bf F}_{[k}}^{\nu \mu }e_{l]\nu}-i\frac{\partial {\bf L}}
{{\partial \phi ^{a}}_{,\mu}}(M_{kl}\phi )^{a}
\end{equation}
with
\begin{equation}
{{\bf F}_{k}}^{\nu \mu }\stackrel{\rm def}{=}\frac{\partial {\bf L}}
{\partial {e^{k}}_{\nu ,\mu }}=\frac{\partial {\bf L}_{G}}
{\partial {e^{k}}_{\nu ,\mu }}=-{{\bf F}_{k}}^{\mu \nu }\;.
\end{equation}
For the infinitesimal affine
coordinate transformation
\begin{equation}
{x' }^{\mu }=x^{\mu }+C^{\mu }+{\Omega^{\mu }}_{\nu }x^{\nu }
\end{equation}
with $C^{\mu }$ and ${\Omega^{\mu }}_{\nu }$ being infinitesimal
constants, the generator $G^{\sharp }(\sigma )$ is given by
\begin{equation}
G^{\sharp }(\sigma )=C^{\mu }P_{\mu }-
\frac{1}{2}{\Omega^{\mu }}_{\nu }{L_{\mu }}^{\nu }
\end{equation}
with
\begin{equation}
P_{\mu } \stackrel{\rm def}{=} \int_{\sigma }
\tilde{\bf P}_{\mu }{}^{ \nu }d\sigma_{\nu }\;, \; \;
{L_{\mu }}^{\nu } \stackrel{\rm def}{=} \int_{\sigma }
{{\bf M}_{\mu }}^{\nu \lambda }d\sigma_{\lambda }\;.
\end{equation}
Here, $\tilde{\bf P}_{\mu }{}^{\nu }$ and
${{{\bf M}}_{\mu }}^{\nu \lambda }$ are defined by
\begin{equation}
\tilde{\bf P}_{\mu }{}^{\nu } \stackrel{\rm def}{=}
\tilde{\bf t}_{\mu }{}^{\nu }
+\tilde{\bf T}_{\mu }{}^{\nu }
\end{equation}
with
\begin{equation}
\tilde{\bf t}_{\mu }{}^{\nu } \stackrel{\rm def}{=}
{\delta_{\mu }}^{\nu }{\bf L}_{G}-{{\bf F}_{k}}^{\lambda \nu }
{e^{k}}_{\lambda ,\mu }\;, \; \;
\tilde{\bf T}_{\mu }{}^{\nu } \stackrel{\rm def}{=}
{\delta_{\mu }}^{\nu }{\bf L}_{M}-\frac{\partial {\bf L}_{M}}
{\partial {\phi^{a}}_{,\nu }}{\phi^{a}}_{,\mu }\;,
\end{equation}
and
\begin{equation}
{{\bf M}_{\mu }}^{\nu \lambda }=2({{\bf F}_{k}}^{\nu \lambda }
{e^{k}}_{\mu }
-x^{\nu }\tilde{\bf P}_{\mu }{}^{\lambda })\;,
\end{equation}
respectively. The quantities $P_{\mu }$ and ${L_{\mu }}^{\nu }$
are the canonical energy-momentum and \lq \lq the extended orbital
angular momentum",$^{5),6)}$ respectively, and
$\tilde{\bf t}_{\mu }{}^{\nu }$ and $\tilde{\bf T}_{\mu }{}^{\nu }$
are energy-momentum densities of the gravitational field and of the
matter field $\varphi $, respectively. We have the relation
\begin{equation}
\tilde{\bf T}_{\mu }{}^{\nu }=
\sqrt{-g}{e^{k}}_{\mu }e^{\nu l}T_{kl}\;.
\end{equation}
The transformation rules of $S_{kl}\, ,P_{\mu }$ and
${L_{\mu }}^{\nu }$ and the differential conservation laws for
${{\bf S}_{kl}}^{\mu }\, ,\tilde{\bf P}_{\mu }{}^{\nu }$ and
${{\bf M}_{\mu }}^{\nu \lambda }$
are obtainable in a usual way, if the total action ${\bf I}$ are
invariant under
\vspace*{0.5 cm}\\
-----------------------------------------------------------------------\\
$^{*)}$ The symbol $[\; \; ]$ denotes the antisymmetrization:
$A_{...[k...l]...} \stackrel{\rm def}{=}
(A_{...k...l...}-A_{...l...k...})/2$.\\

                                                \pagebreak
\noindent the corresponding transformations. In cases which we shall
deal with in the following sections, however, the Lagrangian densities
of the source matters are unknown, and the invariance of the total
actions are not guaranteed. But these laws are obtainable on the basis of
the field equation (2$\cdot $9) and of the invariance of the gravitational
action
\begin{equation}
{\bf I}_{G}\stackrel{\rm def}{=}
\frac{1}{c}\int {\bf L}_{G}dx^{0}dx^{1}dx^{2}\;,
\end{equation}
under the corresponding transformations. We need not explicitly assume
the invariance of the {\em total} action ${\bf I}$.

Under the transformation (2$\cdot $15), $S_{kl}\, ,P_{\mu }$ and
${L_{\mu }}^{\nu }$ transform according as
\begin{eqnarray}
{S'}_{kl}&=&S_{kl}+{d_{k}}^{m}S_{ml}+{d_{l}}^{m}S_{km}\;,\\
{P'}_{\mu }&=&P_{\mu }\;,\\
{{L'}_{\mu }}^{\nu }&=&{L_{\mu }}^{\nu }\;,
\end{eqnarray}
respectively. Also, under the transformation (2$\cdot $20), they
transform according as
\begin{eqnarray}
{S'}_{kl}&=&S_{kl}\;,\\
{P'}_{\mu }&=&P_{\mu }-{\Omega^{\nu }}_{\mu }P_{\nu }\;,\\
{{L'}_{\mu }}^{\nu }&=&{L_{\mu }}^{\nu }
-{\Omega^{\lambda }}_{\mu }{L_{\lambda }}^{\nu }
+{\Omega^{\nu }}_{\lambda }{L_{\mu }}^{\lambda }
-2C^{\nu }P_{\mu }\;,
\end{eqnarray}
respectively. These rules are derivable by the use of
Eqs. (2$\cdot $9), (2$\cdot $17), (2$\cdot $18), (2$\cdot $22),
(2$\cdot $23), (2$\cdot $25) and (2$\cdot $26).

Next, we consider the differential conservation laws. We first consider
${{\bf S}_{kl}}^{\mu }$. For the case with $M_{kl}=0$, as are the cases
discussed in the following sections,
${{\bf S}_{kl}}^{\mu }$ has the expression
\begin{equation}
{{\bf S}_{kl}}^{\mu }=
-2{{\bf F}_{[k}}^{\nu \mu }e_{l]\nu}\;.
\end{equation}
 From the invariance of the action ${\bf I}_{G}$ under the global
Lorentz transformation (2$\cdot $15), the identity
\begin{equation}
\partial_{\mu }{{\bf S}_{kl}}^{\mu }
-2\frac{\delta {\bf L}_{G}}{\delta {e^{[k}}_{\mu }}e_{l]\mu }
\equiv 0
\end{equation}
follows, which, on the use of the field equation $\delta
{\bf L}/\delta {e^{k}}_{\mu }=0$, leads to
\begin{equation}
\partial_{\mu }{{\bf S}_{kl}}^{\mu }=-2\sqrt{-g}T_{[kl]}\;.
\end{equation}
Thus, we have the differential
conservation law
\begin{equation}
\partial_{\mu }{{\bf S}_{kl}}^{\mu }=0\;,
\end{equation}
when $T_{[kl]}=0$.
Now, we consider $\tilde{\bf P}_{\mu }{}^{\nu }$ and
${{\bf M}_{\mu }}^{\nu \lambda }$. From the invariance of the action
${\bf I}_{G}$ under the general coordinate transformations, the
identity
\begin{equation}
\tilde{\bf P}_{\mu }{}^{\nu }\equiv {e^{k}}_{\mu }
\frac{\delta {\bf L}}{\delta {e^{k}}_{\nu }}
+\partial_{\lambda }({{\bf F}_{k}}^{\nu \lambda }{e^{k}}_{\mu })\;,
\end{equation}
follows, from which we can get the differential conservation laws
\begin{eqnarray}
\partial_{\nu }\tilde{\bf P}_{\mu }{}^{\nu }&=&0\;,\\
\partial_{\lambda }{{\bf M}_{\mu }}^{\nu \lambda }&
=&0\;,
\end{eqnarray}
by using the field equation
$\delta {\bf L}/\delta {e^{k}}_{\mu }=0$.

For the differential conservation laws, the following should be
mentioned: For the dreibeins having singularities, it is highly
desirable to confirm these conservation laws by explicit calculations
in which singularities are treated in a proper way, because the
above derivations are formal ones.
\section*{
\begin{center}
\S~3. Generators for Static Circularly Symmetric Solutions of
Gravitational Field Equation
\end{center}
}
\addtocounter{section}{1}
                                        \setcounter{equation}{0}

\mbox{\hspace*{3ex}}We consider static circularly symmetric
gravitational fields produced by point-like static circular bodies
located at the origin $\vec{r} \stackrel {\rm def}{=}
(x^{1},x^{2})=(0,0)$,
assuming that the \lq \lq spin" of
constituent particles of the bodies can be completely neglected:
$M_{kl}=0$.
For these cases, $({e^{k}}_{\mu })$ can be assumed, without loss
of generality, to have a diagonal $\mbox{\rm form}:^{1)}$
\begin{eqnarray}
({e^{k}}_{\mu })=\left[\begin{array}{ccc} A(r) & 0 & 0 \\
                 0 & B(r) & 0 \\
                 0 & 0 & B(r) \end{array}\right]\;
\end{eqnarray}
with $A$ and $B$ being functions of $r \stackrel {\rm def}{=}
|\vec{r}|$, which leads to $a_{klm}\equiv 0$.

The right hand sides of the field equation (2$\cdot $9) can be
expressed in terms of $A$ and of $B,^{1)}$ and there are the
relations $T_{[kl]}=0\;, \;T_{(0)(1)}=0=T_{(0)(2)}$, where, to avoid
the confusion, Latin indices in $T_{kl}$ are enclosed in the
parentheses. Hence, the differential conservation law (2$\cdot $37)
and the relation $\tilde{\bf T}_{0}{}^{\alpha }
=0=\tilde{\bf T}_{\alpha }{}^{0}$ hold.

For the dreibeins having the form (3$\cdot $1), the quantities
${{\bf F}_{k}}^{\mu \nu }$ have the expression
\begin{eqnarray}
\left\{ \begin{array}{ll}\displaystyle{
{{\bf F}_{(0)}}^{0 \alpha }}&
\displaystyle{=-{{\bf F}_{(0)}}^{\alpha 0}
=C(r)x^{\alpha }\;,\;\alpha =1\;,\;2\;,}\\
\displaystyle{
{{\bf F}_{(1)}}^{1 2}}&
\displaystyle{=-{{\bf F}_{(1)}}^{2 1}=D(r)x^{2}\;,}\\
\displaystyle{
{{\bf F}_{(2)}}^{1 2}}&
\displaystyle{=-{{\bf F}_{(2)}}^{2 1}=-D(r)x^{1}\;,}\\
\displaystyle{
{{\bf F}_{(0)}}^{\alpha \beta}}&
\displaystyle{=0\;,\;{{\bf F}_{k}}^{0 \alpha }=
-{{\bf F}_{k}}^{\alpha 0}=0\;,} \\
&\displaystyle{
k=1\;,\;2\;, \;\alpha \;,\;\beta =1\;,\;2}
\end{array}\right.
\end{eqnarray}
with
\begin{eqnarray}
\left\{ \begin{array}{ll}\displaystyle{
C(r)}&\displaystyle{\stackrel {\rm def}{=}
\frac{1}{2r}\left(3c_{1}\frac{d}{dr}\ln \frac{A}{B}
+4c_{2}\frac{d}{dr}\ln AB \right)\;,}\\
\displaystyle{D(r)}&
\displaystyle{\stackrel {\rm def}{=} -\frac{A}{2rB}
\left(3c_{1}\frac{d}{dr}\ln \frac{A}{B}
-4c_{2}\frac{d}{dr}\ln AB \right)\;.}
\end{array}\right.
\end{eqnarray}
The \lq \lq spin" angular momentum density ${{\bf S}_{kl}}^{\mu }$
can be expressed in terms of $B(r)$ and of
${{\bf F}_{k}}^{\mu \nu }$, and it gives the vanishing
\lq \lq spin" angular momentum:$\; S_{kl}=0$.
The energy-momentum density $\tilde{\bf t}_{\mu }{}^{\nu }$
is expressed as
\begin{eqnarray}
\left\{ \begin{array}{ll}\displaystyle{
\tilde{\bf t}_{0}{}^{\nu }}&\displaystyle{={\delta_{0}}^{\nu }A
\left\{\frac{3c_{1}}{4}
\left(\frac{d}{dr}\ln \frac{A}{B}\right)^{2}
+c_{2}\left(\frac{d}{dr}\ln AB\right)^{2}\right\}} \;,\; \; \;
\displaystyle{\tilde{\bf t}_{\alpha }{}^{0}}\displaystyle{=0 \;,}\\
\displaystyle{
\tilde{\bf t}_{\alpha }{}^{\beta }}
& \displaystyle{ = {\delta_{\alpha }}^{\beta }
\left\{\frac{3c_{1}}{4}
\left(\frac{d}{dr}\ln \frac{A}{B}\right)^{2}
+c_{2}\left(\frac{d}{dr}\ln AB\right)^{2}\right\}}\\
& \displaystyle{
-\frac{x^{\alpha }x^{\beta }A}{2r^{2}}
\left[(3c_{1}+4c_{2})\left\{\left(\frac{d}{dr}\ln A\right)^{2}
+\left(\frac{d}{dr}\ln B\right)^{2}\right\}\right.}\\
&\displaystyle{\left. -2(3c_{1}-4c_{2})
\left(\frac{d}{dr}\ln A\right)
\left(\frac{d}{dr}\ln B\right)\right]\;.}
\end{array}\right.
\end{eqnarray}

The density ${{{\bf M}}_{\mu }}^{\nu \lambda }$ takes the form
\begin{eqnarray}
\left\{ \begin{array}{ll}\displaystyle{
{{\bf M}_{0}}^{0 0}}&
\displaystyle{=-2x^{0}\tilde{\bf P}_{0}{}^{0}\;,\; \;
{{\bf M}_{0}}^{0 \alpha }=2A{{\bf F}_{(0)}}^{0 \alpha }\;,}\\
\displaystyle{{{\bf M}_{0}}^{\alpha 0}}&
\displaystyle{=-2A{{\bf F}_{(0)}}^{0 \alpha }
-2x^{\alpha }\tilde{\bf P}_{0}{}^{0}\;, \; \;
\displaystyle{{{{\bf M}}_{0}}^{\alpha \beta}}=0\;,}\\
\displaystyle{{{\bf M}_{\alpha }}^{\beta \gamma }}&
\displaystyle{=2B{{\bf F}_{(\alpha )}}^{\beta \gamma }
-2x^{\beta }\tilde{\bf P}_{\alpha }{}^{\gamma }\;,\; \;
{{{\bf M}}_{\alpha}}^{0 0}=0\;,}\\
\displaystyle{{{{\bf M}}_{\alpha}}^{0 \beta}}&
\displaystyle{=-2x^{0}\tilde{\bf P}_{\alpha}{}^{\beta }\;,\; \;
{{\bf M}_{\alpha}}^{\beta 0}=0\;.}
\end{array}\right.
\end{eqnarray}
In Ref. 1), exact solutions of the field equation (2$\cdot $9) with
point-like sources have been given, which are normalized as
$A(r_{0})=1=B(r_{0})$ for a radius $r=r_{0}$ and classified into
three cases:
\begin{description}
\item[Case 1.]$3c_{1}+4c_{2}\neq 0\; $:
\begin{equation}
A(r)=X(r)Y(r)\;,\; \;
B(r)=[X(r)]^{(1+\Lambda )/(1-\Lambda )}
[Y(r)]^{(1-\Lambda )/(1+\Lambda )}\;,
\end{equation}
where $\Lambda \stackrel{\rm def}{=} \sqrt{-4c_{2}/3c_{1}}\neq 1$,
and
\begin{equation}
X(r) \stackrel{\rm def}{=}
1+\frac{\Lambda -1}{4\Lambda }K_{1}
\ln \left(\frac{r}{r_{0}}\right)\;, \; \; Y(r) \stackrel{\rm def}{=}
1+\frac{\Lambda +1}{4\Lambda }K_{2}
\ln \left(\frac{r}{r_{0}}\right)\;.
\end{equation}
Here, $K_{1}$ and $K_{2}$ are complex constants satisfying the relation
$K_{1}K_{2}-(1-\Lambda )K_{1}-(1+\Lambda )K_{2}=0$.
\item[Case 2A.]$3c_{1}+4c_{2}=0\; $:
\begin{equation}
A(r)=1+a\ln \left(\frac{r}{r_{0}}\right)\;,\; B(r)=\frac{r_{0}}{r}
\end{equation}
with $a$ being a real constant.
\item[Case 2B.]$3c_{1}+4c_{2}=0\; $:
\begin{equation}
A(r)=1\;,\; B(r)={\left(\frac{r}{r_{0}}\right)}^{b}
\end{equation}
with $b$ being a real constant.
\end{description}
The solutions (3$\cdot $8) and (3$\cdot $9) are also solutions of
the three-dimensional Einstein equation, when $3c_{1}=-1/\kappa $.

We examine the generator $G^{\sharp }(\sigma )$ for each of these
solutions. These solutions have singularities at the origin
$\vec{r}=0$, which will be regularized by replacing
$r$ by $\sqrt{r^{2}+\varepsilon^{2}}$
with $\varepsilon $ being an infinitesimal real constant. The following
should be understood: (1)Any $r$ in expressions of
$X\,,Y\,,A$ and $B$ is replaced with $\sqrt{r^{2}+\varepsilon^{2}}$.
(2)The limit $\varepsilon \rightarrow 0$ is taken at
final stages of calculations.
\begin{description}
\item[Case 1.]:
The energy-momentum density $\tilde{\bf t}_{\mu }{}^{\nu }$ for
this case is
\begin{eqnarray}
\left\{ \begin{array}{ll}\displaystyle{
\tilde{\bf t}_{0}{}^{\nu }}&
\displaystyle{=\frac{3c_{1}}{4}
{\delta_{0}}^{\nu }
\frac{K_{1}K_{2}r^{2}}{(r^{2}+\varepsilon^{2})^{2}}\;,
\; \; \nu =0\;,1\;,2\;,}\\
\displaystyle{
\tilde{\bf t}_{\alpha}{}^{0}}&\displaystyle{=0\;,}\\
\displaystyle{
\tilde{\bf t}_{\alpha }{}^{\beta }}&\displaystyle{=
\frac{3c_{1}}{4}{\delta_{\alpha }}^{\beta }
\frac{K_{1}K_{2}r^{2}}{(r^{2}+\varepsilon^{2})^{2}}
-\frac{3c_{1}}{2}x^{\alpha }x^{\beta }\frac{K_{1}K_{2}}
{{(r^{2}+\varepsilon^{2})}^{2}}\;,\; \; \alpha \;,\beta =1\;,2\;.}
\end{array}\right.
\end{eqnarray}
Also, from Eqs. (2$\cdot $9), (2$\cdot $26) and (3$\cdot $6), we can
get
\begin{eqnarray}
\left\{ \begin{array}{ll}\displaystyle{
\tilde{\bf T}_{0}{}^{0}}
&\displaystyle{=\frac{3c_{1}\varepsilon^{2}}{2(r^{2}
+\varepsilon^{2})^{2}}\{(1+\Lambda )K_{1}Y+(1-\Lambda )K_{2}X\}\;,}\\
\displaystyle{
\tilde{\bf T}_{\alpha }{}^{0}}&\displaystyle{=
0=\tilde{\bf T}_{0}{}^{\alpha }\;,}\\
\displaystyle{
\tilde{\bf T}_{\alpha }{}^{\beta }}
&\displaystyle{=-\frac{3c_{1}}{4}
\frac{K_{1}K_{2}\varepsilon^{2}}{(r^{2}+\varepsilon^{2})^{2}}
{\delta_{\alpha }}^{\beta }\;.}
\end{array}\right.
\end{eqnarray}
The first of Eq. (2$\cdot $22), Eqs. (2$\cdot $23), (3$\cdot $10) and
(3$\cdot $11) give
\begin{eqnarray}
\left\{ \begin{array}{ll}\displaystyle{
P_{0}}&\displaystyle{=\frac{3}{2}\pi c_{1}K_{1}K_{2}
\int_{0}^{\infty }\frac{r^{3}dr}
{(r^{2}+\varepsilon^{2})^{2}}}\\
&\displaystyle{+3\pi c_{1}\varepsilon^{2}
\int_{0}^{\infty }
\frac{\{(1+\Lambda )K_{1}Y+(1-\Lambda )K_{2}X\}r}
{(r^{2}+\varepsilon^{2})^{2}}dr\;,}\\
\displaystyle{
P_{\alpha }}&\displaystyle{=0\;,}
\end{array}\right.
\end{eqnarray}
and
\begin{equation}
\int_{C}\tilde{\bf P}_{\mu }{}^{\nu }d\sigma_{\nu }=0\;,
\end{equation}
where $C$ is the cylinder defined in {\bf Appendix A}.
Equations (2$\cdot $39), (3$\cdot $12) and (3$\cdot $13) satisfy
Eq. (A$\cdot $1) and they are
consistent each other, although $P_{0}$ is diverging. \\
The density ${{{\bf M}}_{\mu }}^{\nu \lambda }$ is evaluated by the
use of Eqs. (2$\cdot $23), (3$\cdot $2), (3$\cdot $5), (3$\cdot $10)
and (3$\cdot $11), and we obtain
the following:
\begin{eqnarray}
{L_{\mu }}^{\nu }
&=&-2{\delta_{\mu }}^{0}{\delta^{\nu }}_{0}x^{0}P_{0}\;,\\
\int_{C}{{\bf M}_{\mu }}^{\nu \lambda }d\sigma_{\lambda }
&=&\frac{3}{2}c_{1}
{\delta_{\mu }}^{0}{\delta^{\nu }}_{0}
\int_{C}\{(1+\Lambda )K_{1}Y+(1-\Lambda )K_{2}X\}
\frac{x^{\alpha}d\sigma_{\alpha }}{r^{2}}\;.
\end{eqnarray}
Here, $t \stackrel{\rm def}{=} x^{0}/c$ stands for the time
coordinate of the space-like surface $\sigma $ on which ${L_{0}}^{0}$
is defined. The component ${L_{0}}^{0}$ is not conserved,
although $\partial_{\lambda }{{\bf M}_{0}}^{0 \lambda }=0$. This is not
a contradiction, as is shown below. There is a relation
\begin{eqnarray}
{L_{0}}^{0}({\sigma_{2}}^{R})-{L_{0}}^{0}({\sigma_{1}}^{R})
=\int_{V^{R}}\partial_{\lambda }{{\bf M}_{0}}^{0 \lambda }
dx^{0}dx^{1}dx^{2}
-\int_{C^{R}}{{\bf M}_{0}}^{0 \lambda }d\sigma_{\lambda }\;.
\end{eqnarray}
Here, $C^{R}$ is a cylinder with the spatial radius $R$ between the
the space-like surfaces ${\sigma_{1}}^{R}\;, {\sigma_{2}}^{R}$, and
the $V^{R}$ is the domain enclosed by the surfaces
$C^{R}\;, {\sigma_{1}}^{R}\;, \; {\sigma_{2}}^{R}$.
The limit $R\rightarrow \infty $ of Eq. (3$\cdot $16), although both
sides of which diverge, corresponds to Eq. (A$\cdot $1) applied to the
generator ${L_{0}}^{0}$. We can show
\begin{eqnarray}
\begin{array}{l}\displaystyle{
\lim_{R \rightarrow \infty}\frac{1}{\ln (R/r_{0})}
\{\mbox{\rm the left hand side of Eq. (3$\cdot $16)}\}} \\
\displaystyle{=\lim_{R \rightarrow \infty}\frac{1}{\ln (R/r_{0})}
\{\mbox{\rm the right hand side of Eq. (3$\cdot $16)}\}
=\mbox{\rm finite.}}
\end{array}
\end{eqnarray}
Thus, we have a consistency in this sense.
\item[Case 2A.]:
The energy-momentum density $\tilde{\bf t}_{\mu }{}^{\nu }$ for this
case is
\begin{eqnarray}
\left\{ \begin{array}{ll}\displaystyle{
\tilde{\bf t}_{0}{}^{\nu }}&
\displaystyle{=\frac{3ac_{1}r^{2}}
{(r^{2}+\varepsilon^{2})^{2}}{\delta_{0}}^{\nu }\;,}\\
\displaystyle{
{{\tilde{\bf t}}_{\alpha }}{}^{0}}&\displaystyle{=0\;,}\\
\displaystyle{
\tilde{\bf t}_{\alpha }{}^{\beta }}&\displaystyle{=
\frac{3ac_{1}r^{2}}{(r^{2}+\varepsilon^{2})^{2}}
{\delta_{\alpha }}^{\beta }
-\frac{6ac_{1}x^{\alpha }x^{\beta }}
{(r^{2}+\varepsilon^{2})^{2}}\;.}
\end{array}\right.
\end{eqnarray}
 From Eqs. (2$\cdot $9), (2$\cdot $26) and (3$\cdot $8), we can get
\begin{eqnarray}
\left\{ \begin{array}{ll}\displaystyle{
\tilde{\bf T}_{0}{}^{0}}
&\displaystyle{=\frac{6c_{1}A\varepsilon^{2}}{(r^{2}
+\varepsilon^{2})^{2}}\;,}\\
\displaystyle{
\tilde{\bf T}_{\alpha }{}^{0}}&\displaystyle{=
0=\tilde{\bf T}_{0}{}^{\alpha }\;,}\\
\displaystyle{
\tilde{\bf T}_{\alpha }{}^{\beta }}
&\displaystyle{=-3ac_{1}
\frac{\varepsilon^{2}}{(r^{2}+\varepsilon^{2})^{2}}
{\delta_{\alpha }}^{\beta }\;.}
\end{array}\right.
\end{eqnarray}

We have
\begin{eqnarray}
\left\{ \begin{array}{ll}\displaystyle{
P_{0}}&\displaystyle{=6\pi ac_{1}\int_{0}^{\infty }\frac{r^{3}dr}
{(r^{2}+\varepsilon^{2})^{2}}
+12\pi c_{1}\varepsilon^{2}
\int_{0}^{\infty }
\frac{Ar}{(r^{2}+\varepsilon^{2})^{2}}dr\;,}\\
\displaystyle{
P_{\alpha}}&\displaystyle{=0\;,}
\end{array}\right.
\end{eqnarray}
and
\begin{equation}
\int_{C}\tilde{\bf P}_{\mu }{}^{\nu }d\sigma_{\nu }=0\;.
\end{equation}
Equations (2$\cdot $39), (3$\cdot $20) and (3$\cdot $21) satisfy
Eq. (A$\cdot $1), although $P_{0}$ is not finite. \\
The density ${{{\bf M}}_{\mu }}^{\nu \lambda }$ is evaluated by the
use of Eqs. (2$\cdot $23), (3$\cdot $2), (3$\cdot $5),
(3$\cdot $8), (3$\cdot $18) and (3$\cdot $19), and we obtain
\begin{eqnarray}
{L_{\mu}}^{\nu}&=&-2x^{0}{\delta_{\mu }}^{0}
{\delta^{\nu }}_{0}P_{0}\;,\\
\int_{C}{{\bf M}_{\mu }}^{\nu \lambda }d\sigma_{\lambda }
&=&6c_{1}{\delta_{\mu }}^{0}{\delta^{\nu }}_{0}
\int_{C}\frac{Ax^{\alpha }}
{r^{2}+\varepsilon^{2}}d\sigma_{\alpha }\;.
\end{eqnarray}
The component ${L_{0}}^{0}$ is diverging and not conserved.
Both sides of Eq. (A$\cdot $1) applied to the generator
${L_{0}}^{0}$ diverge, but we have a consistency in the same
sense as for {\bf Case 1.}.
\item[Case 2B.]:
The energy-momentum density $\tilde{\bf t}_{\mu }{}^{\nu }$
is vanishing for this case:
\begin{equation}
\tilde{\bf t}_{\mu }{}^{\nu }=0\;.
\end{equation}
Also, from Eqs. (2$\cdot $9), (2$\cdot $26) and (3$\cdot $9), we can
get
\begin{equation}
\tilde{\bf T}_{\mu }{}^{\nu }=-6bc_{1}
{\delta_{\mu }}^{0}{\delta^{\nu }}_{0}
\frac{\varepsilon^{2}}
{(r^{2}+\varepsilon^{2})^{2}}\;,
\end{equation}
and we have
\begin{equation}
P_{0}=-6\pi bc_{1}\;,\; \; P_{\alpha}=0\;,
\end{equation}
and
\begin{equation}
\int_{C}\tilde{\bf P}_{\mu }{}^{\nu }d\sigma_{\nu }=0\;.
\end{equation}
Equations (2$\cdot $39), (3$\cdot $26) and (3$\cdot $27) satisfy
Eq. (A$\cdot $1).\\
The density ${{\bf M}_{\mu }}^{\nu \lambda }$ is evaluated by the
use of Eqs. (2$\cdot $23), (3$\cdot $2), (3$\cdot $5), (3$\cdot $9),
(3$\cdot $24) and (3$\cdot $25), which leads to
\begin{eqnarray}
{L_{\mu}}^{\nu}&=&12\pi bc_{1}x^{0}
{\delta_{\mu }}^{0}{\delta^{\nu }}_{0}\;,\\
\int_{C}{{\bf M}_{\mu}}^{\nu \lambda }d\sigma_{\lambda }
&=&-6bc_{1}{\delta_{\mu }}^{0}{\delta^{\nu }}_{0}
\int_{C}\frac{x^{\alpha }}
{r^{2}+\varepsilon^{2}}d\sigma_{\alpha }\;.
\end{eqnarray}
The component ${L_{0}}^{0}$ is not conserved, although
the differential conservation law Eq. (2$\cdot $40) holds. This is
not a contradiction, because we have Eq. (3$\cdot $29) and
Eq. (A$\cdot $1) applied to
${L_{0}}^{0}$ is satisfied.
\end{description}

It is worth adding the following:
For every of {\bf Case 1.},
{\bf Case 2A.} and {\bf Case 2B.}, we have the following:
\begin{description}
\item[(a)]That all of the differential
conservation laws (2$\cdot $37), (2$\cdot $39) and (2$\cdot $40)
are actually satisfied is confirmed by explicit calculations in which the
singularities at $\vec{r}=0$ are treated in a proper way by the
regularization procedure employed above.
\item[(b)]The relation
\begin{equation}
\int_{C}{{\bf S}_{kl}}^{\mu }d\sigma_{\mu }=0
\end{equation}
holds, and Eqs. (2$\cdot $37), (3$\cdot $30) and $S_{kl}=0$ satisfy
Eq. (A$\cdot $1).
\item[(c)] The orbital angular momentum
$L_{[\mu \nu ]} \stackrel{\rm def}{=}
(\eta_{\nu \lambda }{L_{\mu }}^{\lambda }-
\eta_{\mu \lambda }{L_{\nu }}^{\lambda })/2$ is conserved and vanishes,
although the ${L_{0}}^{0}$ is not conserved.
\end{description}
                                              \pagebreak
\section*{
\begin{center}
\S~4. Comment on Momentum and Boosts in Three-Dimensional Einstein Gravity
\end{center}
}
\addtocounter{section}{1}
                                        \setcounter{equation}{0}

\mbox{\hspace*{3ex}}In Ref. 2), Deser has claimed that neither
momentum nor boosts are definable for finite energy solution of the
three-dimensional Einstein gravity. But, this claim is wrong, as
is shown below.

He has considred the metric tensor having the expression
\begin{equation}
\; g_{00}=-1\;,\; g_{0 \alpha }=0\;,\; g_{\alpha \beta}
=\phi \delta_{\alpha \beta }\;,\; \alpha \;,\;\beta =1\;,\;2
\end{equation}
with
\begin{equation}
\phi =\prod_{i=1}^{n}|\vec{r}-\vec{r}_{i}|^{-\alpha_{i}}\;,
\;\alpha_{i}=8Gm_{i}/c^{2}\;,
\end{equation}
which describes the gravitational field outside of particles with
masses $m_{i}$ located at $\vec{r}_{i}$.

Our solutions (3$\cdot $8) and (3$\cdot $9) are also solutions of the
three-dimensional Einstein equation, when the relation
$3c_{1}=-1/\kappa $, which we
shall assume to hold from now on, is satisfied.  The solution
(3$\cdot $9) gives a metric which agrees with the metric
(4$\cdot $1), when $n=1\,,\vec{r}_{1}=0\,,b=-4Gm_{1}/c^{2}\,,r_{0}=1$.
 From the discussion for {\bf Case 2B.}, we see that the
energy-momentum and affine coordinate transformation
including also boosts are all definable, which excludes
Deser's claim mentioned above.

The discussion in Ref. 2) is based on the assumption that the
energy-momentum is given by
\begin{eqnarray}
P_{0}&=&-\frac{1}{2\kappa }
\int\sqrt{^{(2)}g}\; \;^{(2)}R(\{\})dx^{1}dx^{2}\;,\\
P_{\alpha }
&=&\int \partial_{\beta }{\pi_{\alpha }}^{\beta }dx^{1}dx^{2}\;,\; \;
\alpha \;,\; \beta =1\;,\;2
\end{eqnarray}
for the coordinate system $\{x^{0}\,, x^{1}\,, x^{2}\}$ for which the
metric takes the form (4$\cdot $1) and by the corresponding primed
expression for the primed coordinate system
$\{x'^{0}\,, x'^{1}\,, x'^{2}\}$
boosted from $\{x^{0}\,, x^{1}\,, x^{2}\}$. Here, we have defined
$^{(2)}g\stackrel{\rm def}{=}\det (g_{\alpha \beta })$, and
$^{(2)}R(\{\})$ is the two-dimensional Riemann-Christoffel
scalar curvature defined by

                                                   \pagebreak
\begin{eqnarray}
^{(2)} R( \{ \} ) \stackrel{\rm def}{=}{ ^{(2)} g^{\alpha \beta } }
\left(\partial_{\gamma }\mbox{\tiny{$\left\{\mbox{\hspace*{-1.5ex}}
\begin{array}{c}\mbox{\scriptsize $\gamma $} \\
\mbox{\scriptsize $\alpha \: \beta $}\end{array}
\mbox{\hspace*{-1.5ex}}\right\}$}}
-\partial_{\beta }\mbox{\tiny{$\left\{\mbox{\hspace*{-1.5ex}}
\begin{array}{c}\mbox{\scriptsize $\gamma $} \\
\mbox{\scriptsize $\alpha \: \gamma $}\end{array}
\mbox{\hspace*{-1.5ex}}\right\}$}}\right. \nonumber \\
\left.
+\mbox{\tiny{$\left\{\mbox{\hspace*{-1.5ex}}\begin{array}{c}
\mbox{\scriptsize $\gamma $} \\
\mbox{\scriptsize $\delta \: \gamma $}\end{array}
\mbox{\hspace*{-1.5ex}}\right\}$}}
\mbox{\tiny{$\left\{\mbox{\hspace*{-1.5ex}}\begin{array}{c}
\mbox{\scriptsize $\delta $} \\
\mbox{\scriptsize $\alpha \: \beta $}\end{array}
\mbox{\hspace*{-1.5ex}}\right\}$}}
-\mbox{\tiny{$\left\{\mbox{\hspace*{-1.5ex}}
\begin{array}{c}\mbox{\scriptsize $\gamma $} \\
\mbox{\scriptsize $\delta \: \beta $}\end{array}
\mbox{\hspace*{-1.5ex}}\right\}$}}
\mbox{\tiny{$\left\{\mbox{\hspace*{-1.5ex}}
\begin{array}{c}\mbox{\scriptsize $\delta $} \\
\mbox{\scriptsize $\alpha \: \gamma $}\end{array}
\mbox{\hspace*{-1.5ex}}\right\}$}}\right)
\end{eqnarray}
with the two-dimensional Christoffel symbol,
\begin{equation}
 \mbox{\tiny{$\left\{\mbox{\hspace*{-1.5ex}}\begin{array}{c}
     \mbox{\scriptsize $\gamma $}\\
     \mbox{\scriptsize $\alpha \: \beta $}
     \end{array}\mbox{\hspace*{-1.5ex}}\right\}$}}
     \stackrel{\rm def}{=} \frac{1}{2}\:{^{(2)}g^{\gamma \delta }}
                           (\partial_{\alpha }g_{\delta \beta }
                           +\partial_{\beta}g_{\delta \alpha }-
                           \partial_{\delta }g_{\alpha \beta})\;,
\end{equation}
where $(^{(2)}g^{\alpha \beta })\stackrel{\rm def}{=}
(g_{\alpha \beta })^{-1}$.
Also, ${\pi_{\alpha }}^{\beta }\stackrel{\rm def}{=}
g_{\alpha \gamma}\pi^{\gamma \beta }$ with $\pi^{\gamma \beta }$ being
the momentum conjugate to $g_{\gamma \beta }$. This momentum has
the expression
\begin{equation}
\pi^{\gamma \beta } =
\frac{\sqrt{^{(2)}{g}}}{2\kappa }
(K^{\gamma \beta }-\; ^{(2)}g^{\gamma \beta }K)
\end{equation}
with
\begin{eqnarray}
K^{\gamma \beta } &\stackrel{\rm def}{=}&
^{(2)}g^{\gamma \delta }\; \, {^{(2)}g^{\beta \epsilon }}
K_{\delta \epsilon }\stackrel{\rm def}{=}\,{
^{(2)}g^{\gamma \delta }}\; \, {^{(2)}g^{\beta \epsilon }}
\left(\frac{\partial_{0}g_{\delta \epsilon }
-\nabla_{\delta }\lambda_{\epsilon }
-\nabla_{\epsilon }\lambda_{\delta }}
{2\sqrt{N}}\right)\;,\\
K &\stackrel{\rm def}{=}&
^{(2)}g^{\alpha \beta }K_{\alpha \beta }\;,\\
\nabla_{\alpha }\lambda_{\beta } &\stackrel{\rm def}{=}&
\partial_{\alpha }\lambda_{\beta }-
 \mbox{\tiny{$\left\{\mbox{\hspace*{-1.5ex}}\begin{array}{c}
     \mbox{\scriptsize $\gamma $}\\
     \mbox{\scriptsize $\alpha \: \beta $}
     \end{array}\mbox{\hspace*{-1.5ex}}\right\}$}}
     \lambda_{\gamma }\;, \\
N &\stackrel{\rm def}{=}& -\frac{1}{g^{00}}\;,\; \;
\lambda_{\alpha } \stackrel{\rm def}{=} g_{0 \alpha }\;.
\end{eqnarray}
Here, in Eqs. (4$\cdot $5)$\sim $(4$\cdot $11), the indices
$\alpha \,,\beta\,, \gamma \,,\delta $ and $\epsilon $ take the
values 1 and 2.

This assumption is groundless. For momentum
of an asymptotically flat generic solution of four-dimensional
Einstein equation, a representation corresponding to Eq. (4$\cdot $4)
holds for any coordinate system related by Lorentz transformations to
an asymptotically Cartesian coordinate system. But, this does not
justify the above assumption for the three-dimensional case.

In the coordinate system $\{x^{0}\,,x^{1}\,,x^{2}\}$, the energy-momentum
$P_{\mu }$, defined as the generator of the space-time translations,
can be expressed as
\begin{eqnarray}
P_{0}&=&\int \tilde{\bf P}_{0}{}^{0}dx^{1}dx^{2}=-\frac{1}{2\kappa }
\int \sqrt{^{(2)}g}\; \; ^{(2)}R(\{\})dx^{1}dx^{2}
=\frac{1}{\kappa }\int \Delta \ln Bdx^{1}dx^{2}\;,\\
P_{\alpha }&=&\int \tilde{\bf P}_{\alpha }{}^{0}dx^{1}dx^{2}=0
=\int \partial_{\beta }{\pi_{\alpha }}^{\beta }dx^{1}dx^{2}\;,
\end{eqnarray}
for the solution (3$\cdot $9), where we have defined
$\Delta \stackrel{\rm def}{=}(\partial_{1})^{2}+(\partial_{2})^{2}$.
In the coordinate system $\{x'^{0},x'^{1},x'^{2}\}$,
however, we have
\begin{eqnarray}
{P'}_{0}&\stackrel{\rm def}{=}&
\int {\tilde{\bf P}'}_{0}{}^{0}dx'^{1}dx'^{2}
\neq -\frac{1}{2\kappa }
\int \sqrt{^{(2)}g'}\; \; ^{(2)}R'(\{\})dx'^{1}dx'^{2}\;,\\
{P'}_{\alpha }&\stackrel{\rm def}{=}&
\int {\tilde{\bf P}'}_{\alpha }{}^{0}dx'^{1}dx'^{2}
\neq \int {\partial'}_{\beta }{{\pi'}_{\alpha }}^{\beta }
dx'^{1}dx'^{2}\;.
\end{eqnarray}
Thus, the discussion in Ref. 2) is wrong in this respect.

The energy-momentum and general affine coordinate transformations
including also boosts are definable for the solution (4$\cdot $1) of
the three-dimensional Einstein equation.
\section*{
\begin{center}
\S~5. Summary and Remarks
\end{center}
}
\addtocounter{section}{1}
                                         \setcounter{equation}{0}

\mbox{\hspace*{3ex}}In the above, we have examined \lq \lq spin"
angular momentum, energy-momentum and the \lq \lq extended orbital
angular momentum" in a teleparallel theory of (2+1)-dimensional
gravity. Also, we have given a critical comment on the discussion in
Ref. 2) claiming that neither momentum nor boosts are definable for
finite energy solution of the three-dimensional Einstein gravity.
The results can be summarized as follows:
\begin{description}
\item[(1)] Under the global Lorentz transformation (2$\cdot $15),
the \lq \lq spin" angular momentum $S_{kl}$, the energy-momentum
$P_{\mu }$ and the \lq \lq extended angular momentum"
${L_{\mu }}^{\nu }$ transform according as Eqs. (2$\cdot $28),
(2$\cdot $29) and (2$\cdot $30), respectively. Also, they transform
according as Eqs. (2$\cdot $31), (2$\cdot $32) and (2$\cdot $33),
under the infinitesimal affine coordinate transformation
Eqs. (2$\cdot $20).
\item[(2)]We have {\em formally} derived the differential
conservation laws (2$\cdot $37), (2$\cdot $39) and (2$\cdot $40),
{\em without assuming explicitly} the invariance of the {\em total}
action {\bf I}.
\item[(3)]For the static circularly symmetric solutions (3$\cdot $6),
(3$\cdot $8) and (3$\cdot $9), we have obtained the following:
\begin{description}
\item[(A)]All the differential conservation laws
(2$\cdot $37), (2$\cdot $39) and (2$\cdot $40) are
{\em actually satisfied} by every of the solutions (3$\cdot $6),
(3$\cdot $8) and (3$\cdot $9), although the dreibeins are
singular at the origin $\vec{r}=0$.
\item[(B)]All the \lq \lq spin" angular momentum $S_{kl}$,
the momentum $P_{\alpha }$ and the orbital angular momentum
$L_{[\mu \nu ]}$ vanish for static circularly symmetric
solutions.
\item[(C)]For each of the solutions (3$\cdot $6) and (3$\cdot $8), the
energy $-P_{0}$ is conserved and divergent. The component
${L_{0}}^{0}$ is diverging and not conserved. But, there arises no
inconsistency.
\item[(D)] For the solution (3$\cdot $9), the energy $-P_{0}$ and the
component ${L_{0}}^{0}$ of the \lq \lq extended orbital angular
momentum" have the finite values $6\pi bc_{1}$ and
$12\pi bc_{1}x^{0}$, respectively. There is no inconsistency, although
${L_{0}}^{0}$ is not conserved.
\end{description}
\item[(4)] Both of the energy-momentum and general affine coordinate
transformations can be defined for the solution (4$\cdot $1), which
invalidates Deser's $\mbox{\rm claim}^{2)}$ that neither momentum nor
boosts are definable for finite energy solutions of three-dimensional
Einstein gravity.
\end{description}

Gravitating bodies for {\bf Case 1.}, {\bf Case 2.A.} and
{\bf Case 2.B.} are all localized at the origin $\vec{r}=0$,
which is known from the expressions (3$\cdot $11), (3$\cdot $19)
and (3$\cdot $25) for the energy-momentum densities
$\tilde{\bf T}_{\mu}{}^{\nu }$ of source bodies by noting
the relation
\begin{equation}
\lim_{\varepsilon \rightarrow 0}\frac{\varepsilon^{2}}
{(r^{2}+\varepsilon^{2})^{2}}=\pi \delta (\vec{r})\;.
\end{equation}
The source body for {\bf Case 2B.} is a mass point with the mass
$6\pi bc_{1}/c^{2}$. We can give {\em tentative} interpretations
to the sources for {\bf Case 1.} and for {\bf Case 2A.}.
When the source of gravity is fluid having the four-velocity field
$u^{\mu }(x)$, the mass density $\rho (x)$ and the pressure
$p(x)$, the density $\tilde{\bf T}_{\mu}{}^{\nu }$ is given by
\begin{equation}
\tilde{\bf T}_{\mu}{}^{\nu }
=\sqrt{-g}\left[\rho (x)u_{\mu }(x)u^{\nu}(x)
+p(x)\left\{\frac{1}{c^{2}}u_{\mu }(x)u^{\nu }(x)
+{\delta_{\mu }}^{\nu }\right\}\right]\;,
\end{equation}
which reduces to
\begin{eqnarray}
\left\{ \begin{array}{ll}\displaystyle{
\tilde{\bf T}_{0}{}^{0}}&\displaystyle{=-\rho (x)c^{2}\sqrt{-g}\;,}\\
\displaystyle{
\tilde{\bf T}_{\alpha }{}^{0}}&\displaystyle{=
0=\tilde{\bf T}_{0}{}^{\alpha }\;,}\\
\displaystyle{
\tilde{\bf T}_{\alpha }{}^{\beta }}&
\displaystyle{=p(x)\sqrt{-g}{\delta_{\alpha }}^{\beta }\;,}
\end{array}\right.
\end{eqnarray}
when the fluid is at rest:
$\sqrt{-g_{00}(x)}u^{0}(x)=-u_{0}(x)/\sqrt{-g_{00}(x)}=c\, ,
u^{\alpha}(x)=0\, ,\alpha =1,2$. Thus, if we {\em make bold to regard}
the gravitating bodies for {\bf Case 1.} and for {\bf Case 2A.} as
fluid, we get the following mass density and pressure for each
case:\\
{\bf Case 1.}$\; $: \\
\begin{eqnarray}
\left\{ \begin{array}{ll}\displaystyle{
\rho (x)}&\displaystyle{=-\frac{3c_{1}\varepsilon^{2}}{2c^{2}(r^{2}
+\varepsilon^{2})^{2}}\{(1+\Lambda )K_{1}Y+(1-\Lambda )K_{2}X\}}\\
&\times \displaystyle{[X(r)]^{(\Lambda+3)/(\Lambda -1)}
[Y(r)]^{(\Lambda -3)/(\Lambda +1)}\;,}\\
\displaystyle{
p(x)}&\displaystyle{=-\frac{3c_{1}}{4}
\frac{K_{1}K_{2}\varepsilon^{2}}{(r^{2}+\varepsilon^{2})^{2}}
[X(r)]^{(\Lambda+3)/(\Lambda -1)}
[Y(r)]^{(\Lambda -3)/(\Lambda +1)}\;.}
\end{array}\right.
\end{eqnarray}
{\bf Case 2A.}$\; $: \\
\begin{eqnarray}
\left\{ \begin{array}{ll}\displaystyle{
\rho (x)}&\displaystyle{=-\frac{6c_{1}\varepsilon^{2}}
{c^{2}{r_{0}}^{2}(r^{2}+\varepsilon^{2})}\;,}\\
\displaystyle{
p(x)}&\displaystyle{=-\frac{3ac_{1}\varepsilon^{2}}
{{r_{0}}^{2}A(r^{2}+\varepsilon^{2})}\;.}
\end{array}\right.
\end{eqnarray}
For {\bf Case 2A.}, the mass density and the pressure both vanish in the
limit $\varepsilon \rightarrow 0$. But, this does not mean that
there is no source of gravity, because $\tilde{\bf T}_{0}{}^{0}$ gives
non-trivial (diverging) contribution to $P_{0}$, as is seen from the
first of Eq. (3$\cdot $20).
\renewcommand{\theequation}{\mbox{A$\cdot$\arabic{equation}}}
\section*{
\begin{center}
Appendix A \\
-----{\em  Relation between Current and Charge}-----
\end{center}
}
                                         \setcounter{equation}{0}

\mbox{\hspace*{3ex}}In general, we have the relation
\begin{equation}
Q_{a}(\sigma_{2})-Q_{a}(\sigma_{1})=\int_{V}\partial_{\mu }
{{\bf j}_{a}}^{\mu }dx^{0}dx^{1}dx^{2}
-\int_{C}{{\bf j}_{a}}^{\mu }d\sigma_{\mu }\;,
\end{equation}
between the current ${{\bf j}_{a}}^{\mu }\, (a=1,2,...,N)$ and
the charges
\begin{equation}
Q_{a}(\sigma_{i})
=\int_{\sigma_{i}}{{\bf j}_{a}}^{\mu }d\sigma_{\mu }\;,
\;i=1\;,\; 2\;.\\
\end{equation}
Here $\sigma_{1}$ and $\sigma_{2}$ are space-like surfaces, $C$
is the cylinder between these surfaces at spatial infinity, and $V$
is the domain of the space-time enclosed by $\sigma_{1}\,, \sigma_{2}$
and $C$. Thus, we have the conservation law
$Q_{a}(\sigma_{2})=Q_{a}(\sigma_{1})\,,$ if
\begin{equation}
\partial_{\mu}{{\bf j}_{a}}^{\mu }=0\;,\; \;
\int_{C}{{\bf j}_{a}}^{\mu }d\sigma_{\mu }=0\;.\\
\end{equation}

\end{document}